# A Subjective and Objective Study of Space-Time Subsampled Video Quality

Dae Yeol Lee, Somdyuti Paul, Christos G. Bampis, Hyunsuk Ko, Jongho Kim, Se Yoon Jeong,
Blake Homan, and Alan C. Bovik

*Abstract*—**Video dimensions are continuously increasing to provide more realistic and immersive experiences to global streaming and social media viewers. However, increments in video parameters such as spatial resolution and frame rate are inevitably associated with larger data volumes. Transmitting increasingly voluminous videos through limited bandwidth networks in a perceptually optimal way is a current challenge affecting billions of viewers. One recent practice adopted by video service providers is space-time resolution adaptation in conjunction with video compression. Consequently, it is important to understand how different levels of space-time subsampling and compression affect the perceptual quality of videos. Towards making progress in this direction, we constructed a large new resource, called the ETRI-LIVE Space-Time Subsampled Video Quality (ETRI-LIVE STSVQ) database, containing 437 videos generated by applying various levels of combined space-time subsampling and video compression on 15 diverse video contents. We also conducted a large-scale human study on the new dataset, collecting about 15,000 subjective judgments of video quality. We provide a rate-distortion analysis of the collected subjective scores, enabling us to investigate the perceptual impact of space-time subsampling at different bit rates. We also evaluated and compare the performance of leading video quality models on the new database.**

## I. INTRODUCTION

THE streaming and social media industry is continuously progressing towards providing more realistic and immersive experiences to video consumers. Display companies and content providers are enabling higher spatial resolutions, frame rates, and high dynamic range (HDR). Televisions and monitors are now available that support 8K HDR and/or true 120Hz 10-bit input and playout. Popular media streaming services, such as YouTube, Netflix, and Amazon, now provide

D. Lee, S. Paul, and A.C. Bovik are with the Department of Electrical and Computer Engineering, The University of Texas at Austin, Austin, TX, USA (email:daelee711@utexas.edu, somdyuti@utexas.edu, bovik@ece.utexas.edu). C.G.Bampis is with Netflix Inc., Los Gatos, CA, USA (email: christosb@netflix.com). H. Ko is with Division of Electrical Engineering, Hanyang University ERICA, Ansan, South Korea (email: hyunsuk@hanyang.ac.kr). J. Kim and S. Jeong are with Realistic AV Research Group, ETRI, Daejeon, South Korea (email: pooney@etri.re.kr, jsy@etri.re.kr). B. Homan is the CEO of Video Clarity, Inc., Campbell, CA, USA (blake@videoclarity.com).

contents at 4K/60fps/HDR, and it is expected that increases in these video parameters will be met by even larger, faster, and deeper displays and streamed video content. However, increases in video dimensions inevitably increase streamed data volume, hence service providers are increasingly challenged to deliver high-quality videos with limited bandwidths. while providing the highest possible quality.

Video compression is the principal technology that enables bandwidth-constrained video streaming, as exemplified by the global ITU standards H.264 [1], HEVC [2], and the emerging Versatile Video Coder (VVC), as well as the open source standards VP9 [3] and AV-1 [4].

Given increases in video dimensions, a recent approach taken by streaming video providers is to combine resolution adaptation with compression. For example, a spatially subsampled video may require less quantization (compression) to meet a given bit rate requirement, and possibly resulting in a perceptually less degraded video, depending on the content. Thus far, this practice has been largely limited to spatial subsampling, but temporal, and more generally, space-time subsampling, also offer the potential for increased efficiencies.

There have been some studies that have investigated the combined effects of spatial subsampling and compression on perceptual video quality [5], [6]. The authors of [7] investigated the perceptual quality of videos with varying spatial adaptation filters including the nearest-neighbor, bicubic, and Convolutional Neural Network (CNN) based super resolution filter.

Other authors have studied temporal subsampling and its effects on subjective video quality, but without considering coincident compression or spatial subsampling. They used these results to motivate resolution adaptation methods which reduce video frame rate if the content does not perceptually benefit from a higher frame rate [8], [9]. In [10], a spatio-temporal resolution adaptation method for video compression was proposed, but quality prediction and consequent downsampling decisions were conducted separately in space and time. The authors of [11] did study the joint application of space-time subsampling and compression, but the codec was confined to H.264, and the maximum considered frame rate was 60 Hz.

These prior efforts have helped us to understand how spatial and temporal video density affect perceptual quality. However,

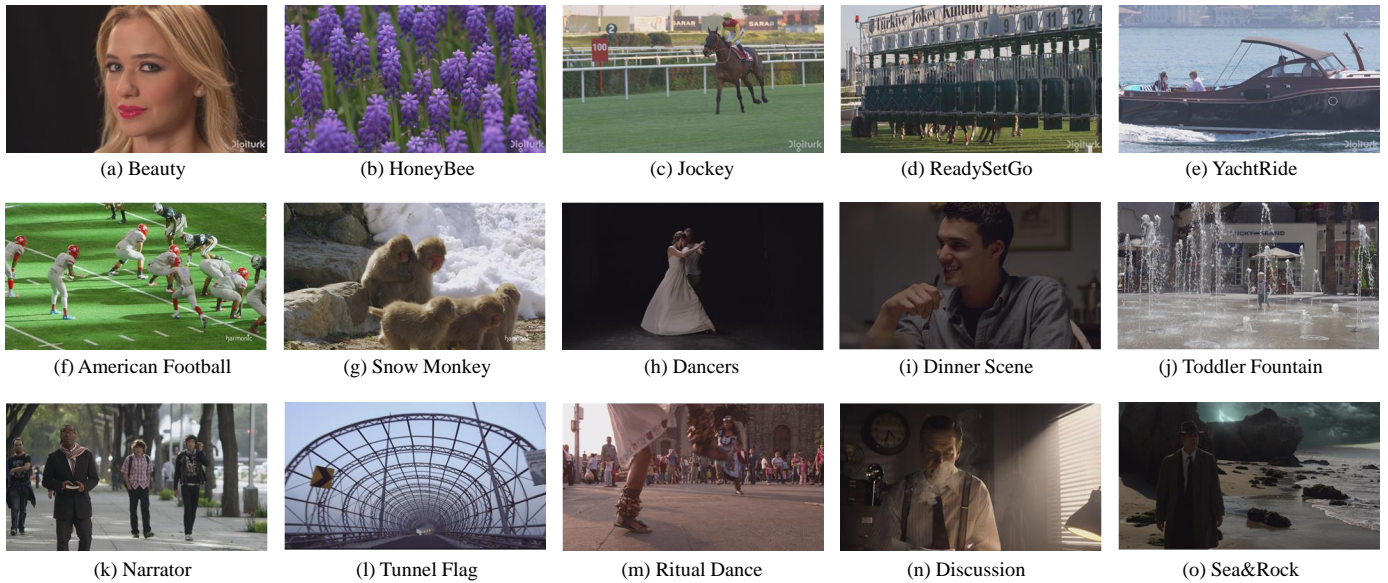

Fig. 1. Sample frames from source contents in the ETRI-LIVE Space-Time Subsampled Video Quality Database

TABLE I
SUMMARY OF SOURCE CONTENT VIDEO FORMATS

| Video | Bit Depth | Chroma Format | | Spatial Resolution | | Frame rate (fps) | No. of Frames | Duration (sec) |
|---|---|---|---|---|---|---|---|---|
| | | Original | Processed | Original | Processed | | | |
| Beauty | 10 | YUV420p | YUV420p | 3840×2160 | 3840×2160 | 120 | 600 | 5 |
| HoneyBee | 10 | YUV420p | YUV420p | 3840×2160 | 3840×2160 | 120 | 600 | 5 |
| Jockey | 10 | YUV420p | YUV420p | 3840×2160 | 3840×2160 | 120 | 600 | 5 |
| ReadySetGo | 10 | YUV420p | YUV420p | 3840×2160 | 3840×2160 | 120 | 600 | 5 |
| YachtRide | 10 | YUV420p | YUV420p | 3840×2160 | 3840×2160 | 120 | 600 | 5 |
| Snow Monkey | 10 | YUV420p | YUV420p | 3840×2160 | 3840×2160 | 60 | 360 | 6 |
| American Football | 10 | YUV420p | YUV420p | 3840×2160 | 3840×2160 | 60 | 420 | 7 |
| Dancers | 10 | YUV420p | YUV420p | 4096×2160 | 3840×2160 | 60 | 404 | 6.7 |
| Dinner Scene | 10 | YUV420p | YUV420p | 4096×2160 | 3840×2160 | 60 | 360 | 6 |
| Toddler Fountain | 10 | YUV420p | YUV420p | 4096×2160 | 3840×2160 | 60 | 420 | 7 |
| Narrator | 10 | YUV420p | YUV420p | 4096×2160 | 3840×2160 | 60 | 300 | 5 |
| Tunnel Flag | 10 | YUV420p | YUV420p | 4096×2160 | 3840×2160 | 60 | 360 | 6 |
| Ritual Dance | 10 | YUV420p | YUV420p | 4096×2160 | 3840×2160 | 60 | 280 | 4.7 |
| Discussion | 10 | YUV422p | YUV420p | 3840×2160 | 3840×2160 | 60 | 380 | 6.3 |
| Sea&Rock | 10 | YUV422p | YUV420p | 3840×2160 | 3840×2160 | 60 | 272 | 4.5 |

given that spatial and temporal (space-time) subsampling and compression are likely to be applied in concert, studies are needed to be able to understand and model how they affect perceived video quality when they are jointly applied. Towards advancing progress in this direction, we have constructed a large-scale video quality database entitled the "ETRI-LIVE Space-Time Subsampled Video Quality (ETRI-LIVE-STSVQ)" database, which contains a large number of videos operating at different space resolutions, temporal frame rates, and levels of compression, along with collected subjective human opinion scores on all of them. The contributions that we make include:

- The first database with subjective quality scores rendered on 4K 10-bit videos at frame rates up to 120Hz, subjected to simultaneous space-time subsampling and compression (HEVC) distortions applied at multiple levels. A total of 437 space-time subsampled and compressed videos were created.

- We conducted a large-scale laboratory human study on

the videos, using a high-speed video playout system capable of displaying true 120 Hz 10-bit video signals in real-time.

- Since the new database can be uniquely used to design and compare video quality models that can predict the perceptual quality of space-time subsampled and compressed videos, we conducted a comparative study of relevant popular VQA models on the prediction problem.

- The new database is a unique psychometric resource for understanding the perceptual effects of space-time subsampling and compression, and for designing strategies for subsampling and compression parameter control to achieve perceptually optimized target bitrates.

The rest of the paper is organized as follows: Section II provides a detailed description of the construction of the database. Section III describes the subjective experiment protocol. Section IV describes the data processing and analysis of the subjective opinion scores. Section V compares the performances of various relevant high-performance video

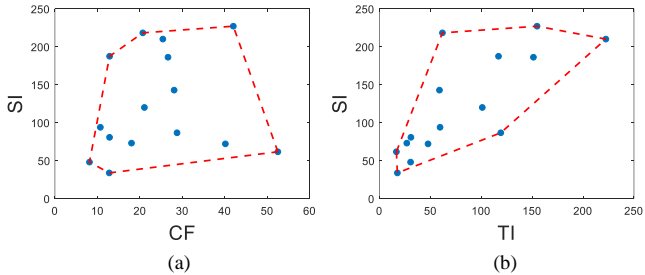

Fig. 2. Spatial Information (SI) versus colorfulness (CF), and (b) Temporal Information (TI) versus colorfulness (CF) measured on the source contents of the ETRI-LIVE STSVQ database. The convex hull is indicated in red boundaries.

quality models on the new database. Finally, conclusions are drawn in Section VI.

## II. CONSTRUCTION OF THE DATABASE

### A. Source Contents

We collected 15 high quality 4K 10-bit source contents having wide variety of spatiotemporal properties. Of the 15 contents, five are from the Ultra Video Group (UVG) dataset [12], two are Harmonic 4K footages, and eight are from the Netflix public video library [13]. Fig. 1 shows sample frames of each of the source contents, while Table I the formats of the source videos. As shown in the table, all of the source contents are of high spatial resolutions and frame rates of at least 3840×2160 and 60 fps, respectively. We set the target video format of the source contents as 3840×2160, YUV420p, and 10 bits. A few contents of spatial resolution 4096×2160 were slightly cropped, of format YUV422p were chroma subsampled, to meet the target video format. Among the 15 source contents, five taken from the UVG dataset have frame rates of 120fps, while the other ten have frame rates of 60fps. Each video content was clipped to the range 5~7 second duration, taking care to exclude scene changes or disruptions of content, such as a sport play or an actor speaking. The average duration of the video contents is 5.61 seconds.

The diversity of the source contents was confirmed by measuring the spans of (i) low-level space-time video features and (ii) encoding complexities. The low-level feature measurements included the spatial information (SI) and temporal information (TI) suggested in [14], representing the complexity of spatial details and temporal change of the videos, respectively. Another low-level feature that was used is the colorfulness (CF) measure proposed in [15]. Figs. 2(a) and (b) show plots of SI against CF and SI against TI with their corresponding convex hulls superimposed. The plots illustrate a diverse span of spatial and temporal characteristics covered by the source contents. We also computed the relative range and the uniformity of coverage [16] on each low-level feature, to quantify how well the feature space is covered by the selected source contents. The relative ranges of SI, TI, and CF were 0.85, 0.93, and 0.85, respectively, and the uniformity of coverage values for SI, TI, and CF were 0.84, 0.80, and 0.81, respectively. Again, the values illustrate the diverse space-time characteristics of the selected contents. We also considered content complexity as measured by encoded bitrate [17]. We encoded all of the source contents using HEVC (libx265) with a fixed quantization parameter (QP) of 29, then measured the bit rate of each content. As shown in Fig. 3, the source contents span a wide range of encoding complexities, ranging from less than 1Mbps to 130Mbps.

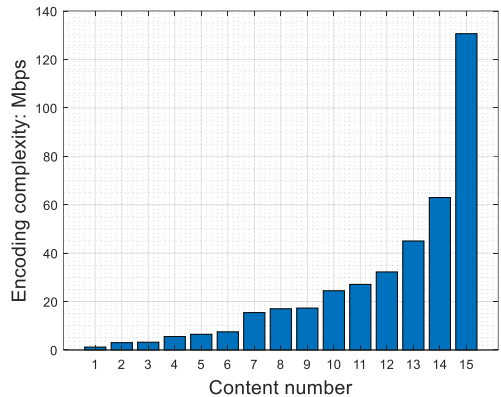

Fig. 3. Encoding complexity across contents, expressed in terms of Mbps when encoding using HEVC (libx265) at QP 29.

### B. Distorted Video Generation

Each of the 15 source contents was subjected to various levels of distortion, in the form of space-time subsampling and compression. Since a main goal of our study is to understand the joint effects of space-time subsampling and compression, with an aim to improve perceptually optimal video coding strategies in practical settings, we constrained the videos used in the experiments to each approximate one of five target bit rates. These cover a range of perceived video qualities from very high to very low, while allowing for noticeable perceptual separations between bit rate levels. We then generated distorted videos having various combinations of space-time subsampling and degree of compression to approximately meet the predefined target bit rates. In this way, we generated 437 distorted videos affected by space-time subsampling and compression.

In the subjective study to be described shortly, all of the videos that were rated were viewed on a display supporting the target video format of 3840×2160, 60/120 fps, YUV420p, and 10 bits. Hence, subsampled videos were up-sampled back to the target format before being viewed. Fig. 4 shows the processing flow on space-time subsampled videos for our subjective experiment. As shown in the figure, space-time subsampled videos are restored to the target space-time resolutions before being viewed, thereby avoiding visual effects by the display's space-time up-sampling engines. Next, we explain how each distortion was applied to the source videos.

#### 1) Spatial subsampling

The database includes videos of four different spatial resolutions, including the source resolution (3840×2160) and three subsampled resolution (1920×1080, 1280×720, and 960×540). The videos were down-sampled prior to encoding using the Lanczos kernel [18]. The spatially subsampled videos were then up-sampled back to the target spatial resolution

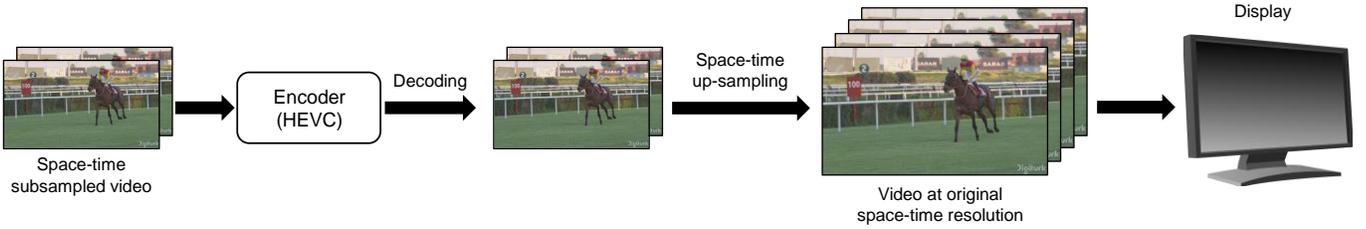

Fig. 4. Workflow for viewing space-time subsampled videos in the subjective experiment described in Section III.

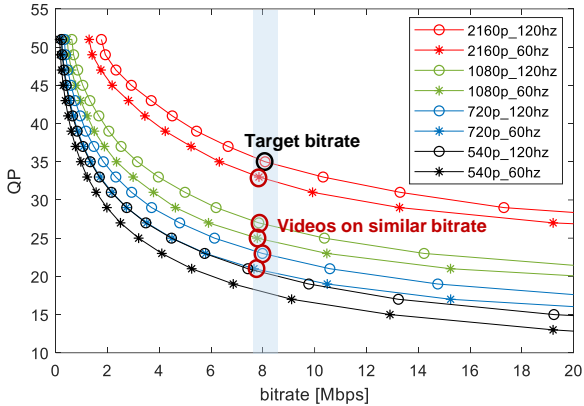

Fig. 5. Example of QP determination on space-time subsampled versions of the 'Jockey' video.

(3840×2160), also using the Lanczos kernel, prior to displaying them.

### 2) Temporal subsampling

The database contains original source videos that have frame rates of 120 or 60 fps, and temporally downsampled ("half frame rate") versions of them at 60 or 30 fps, respectively. The "full frame rate" videos were temporally downsampled to half frame rate, by simply dropping alternate frames, analogous to capturing the video at a lower shutter speed [19], without introducing motion blur.

However, when upsampling videos for viewing by the human subjects (Fig.4), we did not apply simple frame duplication, since this tends to produce visually unpleasant stuttering effects. We also did not rely on the frame rate interpolation engine of the display, since although it is designed to promote motion smoothness, it can produce severe and unexpected distortions. While Motion Compensated Interpolation (MCI) methods can deliver results having high visual quality, they can also introduce severe quality degradations when motion estimation failures occur [20], and they are major contributor to the "soap opera effect". Moreover, there are no agreed-upon best methods of MCI, which varies across display manufacturers. To avoid the severe distortions that can be produced by frame duplication or MCI, we instead utilized Linear Filter Interpolation [21], which linearly interpolates between adjacent frames. Generally, LFI yields stable and consistent results that may be regarded as a lower bound of the best results provided by modern displays, without producing the more severe artifacts that can occur.

### 3) Video compression

The videos were compressed by the Main 10 profile of HEVC, using the FFmpeg libx265 encoder. We fixed the intra period to 1 second, to ensure that I-pictures would be regularly inserted as in the Random Access (RA) configuration of the reference software [22]. The compression levels were controlled by varying the QP parameters, where higher values of QP increased the degrees of compression.

Since the source videos exhibit different space-time characteristics, and consequently, the bitrates arrived at by processing each content with spatial and temporal downsampling and compression were varied by content to span a wide range of perceptual qualities. Instead of imposing the same target bit rates on all contents, we adaptively determined a set of five target bit rates for each source video so that a wide range of perceptual qualities were represented, with good perceptual separations bitrates, and where the space-time subsampled videos were compressed (via QP selection) to have bit rates similar as possible to the target bit rates.

Fig. 5 shows an example of the QP determination process on space-time subsampled versions of the 'Jockey' sequence. The top curve indicates the source content having full space-time resolution, while the lower curves indicate various combinations of space-time subsampling. Once a target bit rate is selected, the QP values of each space-time subsampled video was determined to be most similar to the target bitrate.

## III. SUBJECTIVE EXPERIMENTS

### A. Experiment Design

In subjective experiments, we adopted a Single-Stimulus Continuous Quality Evaluation (SSCQE) procedure with hidden reference [23]. The participants delivered the subjective quality scores using a continuous scale score bar after viewing the video once. The original reference videos are presented but 'hidden', i.e., without being identified as such. The scores on reference videos are useful as high-quality anchors, and are used to calculate difference mean opinion scores (DMOS) as a way of removing content biases.

Given that the average duration of each presented video is 5.61 seconds, and the average time subjects expend scoring each video is about 6 seconds, a participant requires approximately 90 minutes to view and score all of the 437 videos in the database. To avoid viewer fatigue, we therefore divided the study into three 30-minute sessions, each comprising 145 or 146 distorted videos and 15 hidden references. The subjects each participated in three sessions separated by least 24 hours, hence every subject evaluated all of the videos in the database.

When constructing the playlist of videos for each session, we sought to eliminate any biases introduced if the videos were displayed in the same order to every subject. We also avoided the successive presentations of same contents. Therefore, to minimize contextual and memory effects, which can affect judgements of video quality [17], we randomized the playlist using the following procedures.

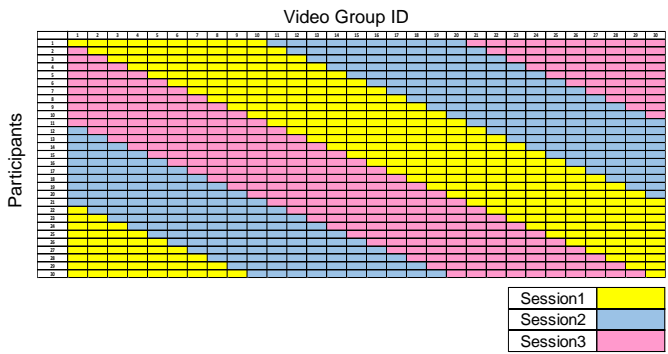

Fig. 6. Illustration of the round-robin method used to determine which video groups are presented to a given participant.

*1) Initial randomization*

An initial randomization was used to reduce any clustering of the source contents. The list was shuffled repeatedly until there were less than 10 occurrences of any video content (distorted or not) that appeared consecutively. Of course, this does not adequately remove the effects of memory, so additional randomization was applied in the final step (to follow).

*2) Video groups*

The randomized 437 videos from the previous steps were then divided into 30 'video groups', each containing 14 or 15 videos. Since each subject participated in three sessions, 10 video groups were viewed in each session.

*3) Round-robin ordering*

We also employed a round-robin presentation ordering, to minimize the possibility of subjective opinions being affected by contextual factors, such as combinations of videos being shown together to all of the participants. Fig. 6 illustrates how the round-robin method was applied to decide which video groups were presented to each participant within each session. As shown in the figure, the video groups comprising sessions 1, 2, and 3 will be different for all participants. This round-robin approach also guarantees each video group will appear an equal number of times within each session (across all subjects). For example, in a study with 30 subjects, the videos in group 1 will appear 10 times in each of sessions 1, 2 and 3.

*4) Final randomization*

After the 10 video groups were selected for a given participant in a given session, the 15 undistorted hidden references were included, and the entire collection of references and 10 video groups were collectively randomly shuffled again into a single session playlist. However, during the shuffling, videos having the same content were constrained to have at least three different contents lie between them, i.e., to be separated by at least four display periods. Proceeding in exactly this way throughout, we generated distinct playlists for every session created throughout the study.

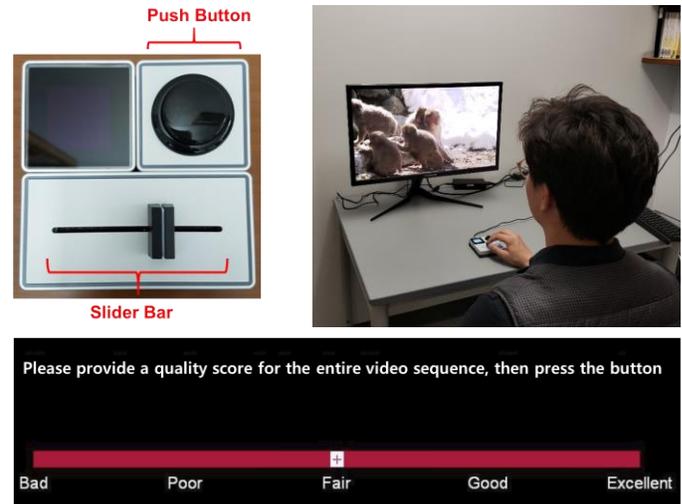

Fig. 7. Hardware slider interface (top) and video score voting screen (bottom).

### B. Experimental set-up

As explained previously, all of the videos included in the study are stored in 10-bit raw YUV format. To be able to successfully play out the UHD high frame rate videos without hitches, we relied on a powerful ClearView system provided by Video Clarity [24], which enables real-time playout of raw 4K 10-bit YUV format videos at 120fps. The system is connected to a 27-inch Acer Predator X27 display which also supports true 120fps 10-bit video signal input and playout [25]. Before each subject entered the study room, the entire experimental dataset and script was preloaded to present the session using the playlist generated for that session. After viewing each test video, a voting screen appeared and the subject used a Pallete hardware slider [26] to control and select the scores from an onscreen slider bar, as shown in Fig. 7. The quality bar is marked with five Likert labels ranging from 'Bad' to 'Excellent.' However, the quality scale is continuous, and the subjects were instructed that they could move the slider bar to any position between the labels. Once the score was selected, it was converted to a numerical value ranging from 0 to 39, where 0 is 'Bad' and 39 is 'Excellent.'

### C. Experiment procedure

Each subject was presented with brief oral summary of the overall experiment, and given written instructions explaining how to use the hardware slider to assign scores to each video, and that scores should reflect the degree of satisfaction they felt regarding the level of overall video quality, while discounting the aesthetic value or interestingness of the content. Each subject was seated in front of the display at a distance of about 1.5 times the height of the display, as recommended in [27] for 4K videos. The subjects then participated in a short training session using 10 videos different from those viewed in the actual study, but also covering a wide range of perceptual qualities and distortions representation of those seen during the actual experiment. The training session enabled the subjects to become familiar with the types and qualities of videos to be judged, to attain facility with the hardware interface used to score them. Following the training session, each subject proceeded immediately to the actual session where the subjective data was collected. The training session was only given prior to each subject's first session.

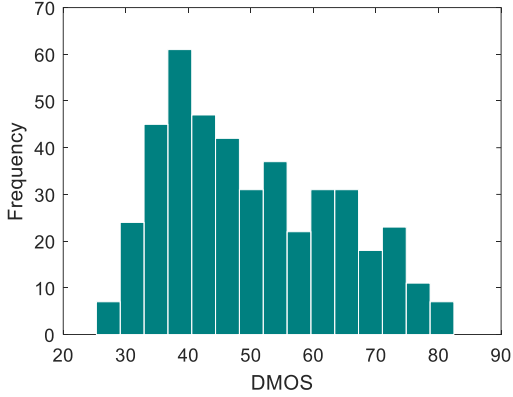

Fig. 8. Histogram of DMOS in 15 equally spaced bins.

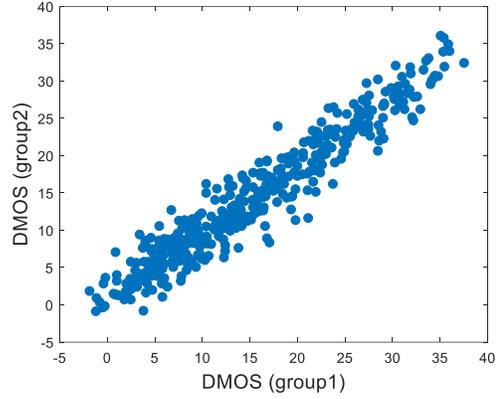

Fig. 9. Scatter plot of DMOS of a random division of the human subjects into two non-overlapping groups of equal size. The plot indicates a strong agreement between the subjects.

## IV. DATA PROCESSING AND ANALYSIS

### A. Processing of Subjective Scores

A total of 34 naïve subjects from The University of Texas at Austin took part in the study. Each subject participated in all three sessions, and thus, each video was rated by all 34 subjects. The collected scores were converted to DMOS according to [28]. Let $s_{ijk}$ refer to the score given by subject $i$ to video $j$ during session $k \in \{1, 2, 3\}$. Then, the difference score $d_{ijk}$ for video $j$ at session $k$ is computed by subtracting the score given to the video from that given to the corresponding (same content) reference video $j_{ref}$ *in the same session $k$*

$$d_{ijk} = s_{ij_{ref}k} - s_{ijk}. \tag{1}$$

The difference scores for the reference videos are, of course, 0, and were then removed from all sessions. To normalize the scores collected on different sessions, we computed the Z-score per session [29] as:

$$\mu_{ik} = \frac{1}{N_{ik}} \sum_{j=1}^{N_{ik}} d_{ijk}, \tag{2}$$

$$\sigma_{ik} = \sqrt{\frac{1}{N_{ik} - 1} \sum_{j=1}^{N_{ik}} (d_{ijk} - \mu_{ijk})^2}, \tag{3}$$

and

$$z_{ijk} = \frac{d_{ijk} - \mu_{ijk}}{\sigma_{ik}}, \tag{4}$$

where $N_{ik}$ is the number of videos the subject $i$ viewed during session $k$. We collected the Z-scores from all sessions and formed a matrix $\{z_{ij}\}$ with element $z_{ij}$ corresponding to the Z-score assigned by subject $i$ to video $j$, where $j \in \{1, 2, ..., 437\}$.

Subject rejection was performed according to the procedure from [23]. The normality of the Z-scores for each content was evaluated by computing the kurtosis $\beta_2$ via

$$\bar{z}_j = \frac{1}{N} \sum_{i=1}^{N} z_{ij}, \tag{5}$$

$$m_{xj} = \frac{\sum_{i=1}^{N} (z_{ij} - \bar{z}_j)^x}{N}, \tag{6}$$

and

$$\beta_{2j} = \frac{m_{4j}}{(m_{2j})^2}, \tag{7}$$

where $N$ refers to the number of subjects that evaluated video $j$, which in our case, is 34. If $2 \le \beta_{2j} \le 4$, we considered the scores for the video $j$ to be normally distributed. We identified potential outlier subjects according to the predicate if $z_{ij} \ge \bar{z}_j + 2\sigma_j$, then $P_i = P_i + 1$, if $z_{ij} \le \bar{z}_j - 2\sigma_j$, then $Q_i = Q_i + 1$,

where

$$\sigma_j = \sqrt{\sum_{i=1}^{N} \frac{(z_{ij} - \bar{z}_j)^2}{N - 1}}. \tag{8}$$

If $\beta_{2j}$ did not fall between 2 and 4, we instead used if $z_{ij} \ge \bar{z}_j + \sqrt{20}\sigma_j$, then $P_i = P_i + 1$,

if $z_{ij} \le \bar{z}_j - \sqrt{20}\sigma_j$, then $Q_i = Q_i + 1$.

In either case, for each subject $i$, we determined if the following conditions:

$$\frac{P_i + Q_i}{N} > 0.05, \tag{9}$$

and

$$\left| \frac{P_i - Q_i}{P_i + Q_i} \right| < 0.3 \tag{10}$$

were met. For a certain subject $i$, if both (9) and (10) were true, then we rejected the subject. Overall, only four out of the 34 subjects were rejected. We linearly rescaled the Z-scores of the remaining 30 subjects to final DMOS values in the range [0, 100] using

$$z'_{ij} = \frac{100(z_{ij} + 3)}{6}. \tag{11}$$

Fig. 8 shows the histogram of the resulting DMOS values,

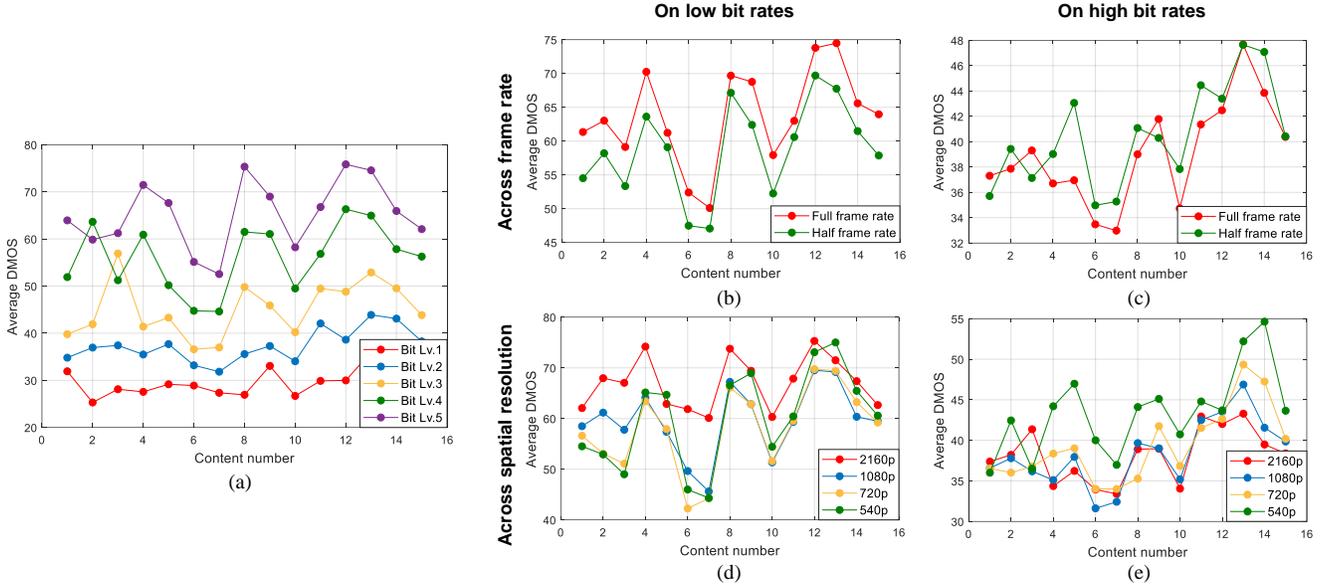

Fig. 10. Variation of average DMOS of each content for each (a) compression level, (b) temporal subsampling level at low bit rates, (c) temporal subsampling level at high bit rates, (d) spatial subsampling level at low bit rates, and (e) spatial subsampling level at high bit rates.

showing a distribution of opinions. The extreme DMOS values were 25.30 and 82.45, while the mean and standard deviation of DMOS were found to be 50 and 13.62, respectively.

To check the consistency of the collected subjective data, we randomly split the subjects into two randomly selected, non-overlapping groups of equal size, and measured the Spearman's Rank Correlation Coefficient (SRCC) between their scores. Fig. 9 shows a scatter plot of DMOS between a pair of such randomly split groups, exhibiting an approximately linear unit slope. The SRCC between the subject halves for this split was 0.958, indicating a high consistency between the groups. This random procedure was repeated 1000 times, yielding SRCC values lying between 0.941 and 0.972 with a median value of 0.960, indicating a reliably high degree of inter-subject consistency.

### B. Analysis of Opinion Scores

We observed how the average DMOS values varied with content across various levels of distortions. Fig. 10(a) plots the average DMOS of each content across the target bit rates. The labels denoted Lv. refer to distortion levels, which in Fig. 10(a) refers to H.265 compression. More specifically, Lv. 1 refers to videos that were compressed to the highest target bit rate, while Lv. 5 indicates the highest degree of compression (lowest target bit rate). The target bit rate decreases with increases of the compression level. Generally, DMOS increased with increased compression, although, as expected, the relationship is not monotonic because of content effects (e.g., masking).

When observing the perceptual effects of different levels of spatial and/or temporal subsampling, consideration must be given to an assumed available bit rate. This is because the spatial and/or temporal subsampling can drive the perceptual quality of a video in different directions depending on an imposed bit budget. For example, videos subjected to little or no space-time subsampling may yield very high levels of perceived quality given a large enough budget. However, the quality may be severely degraded if the budget is small (hence compression is heavy). Figs. 10(b)-(e) depict the effects of different levels of spatial and temporal subsampling of the videos, conditioned on low and high bit rates, where the low bit rates were taken to be levels 4 and 5, and levels 1-3 were regarded as high bit rates.

Figs. 10(b) and (c) plots the average DMOS across full and half frame rates. Fig. 10(c) shows that, at sufficiently high bit rates, the full frame rate videos, generally, yielded lower DMOS (higher perceptual quality) probably in large part because of smoother motion. However, as shown in Fig. 10(b), when the available bit budget was low, the tendency was reversed, and the half frame rate videos resulted in better reported quality than the full frame rate videos. Figs. 10(d) and (e) plot the DMOS across different spatial resolutions. A similar trend may be observed, e.g., in Fig. 10(e), low perceptual quality (higher DMOS) was reported on heavily subsampled 540p videos. However, as shown in Fig. 10(d), at low bit rates, the tendency again reversed, and the 540p videos generally provided better perceptual qualities as compared to the full resolution (2160p) videos. These interesting results nicely exemplify the somewhat complex relationships between spatial and temporal resolution, compression, and perceived quality, and provide evidence that it should be possible to perceptually optimize video coding strategies by considering spatial and/or temporal subsampling combined with compression, especially when trying to attain lower bit rates.

Fig. 11 plots rate distortion (RD) curves on the entire ETRI-LIVE database, further revealing the effects of space-time subsampling. The bit rate and DMOS values of each point on a curve corresponds to the average DMOS of all videos in the database having the same space-time subsampling configuration, as specified in the legend. Note that the vertical

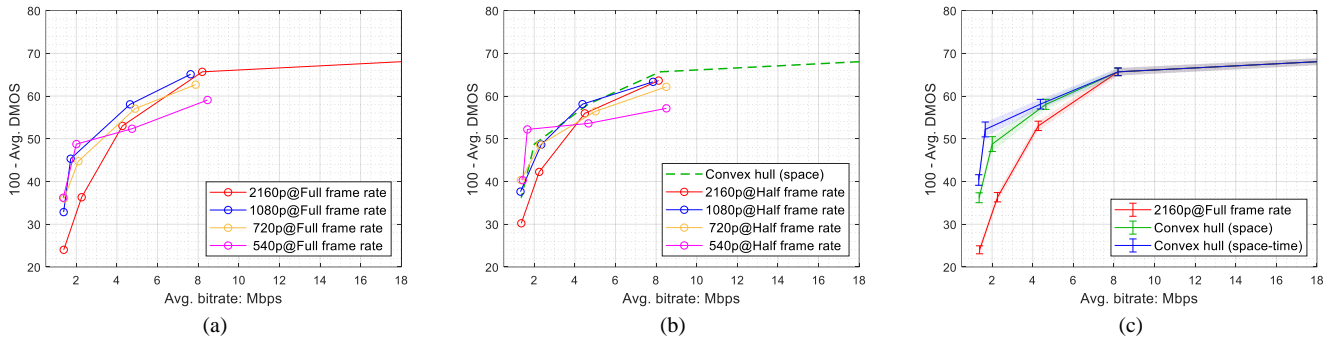

Fig. 11. Rate distortion curves plotted on the entire database to observe the effects of (a) spatial subsampling only, and (b) space-time subsampling. Plot (c) depicts the rate distortion curves of the video at the original space-time resolutions, using the convex hull constructed via spatial subsampling, and the convex hull constructed via space-time subsampling. The error bars represent 95% confidence intervals.

axis is set $100 - Avg. DMOS$, hence higher values correspond to higher perceptual quality.

Fig. 11(a) focuses on the effects of spatial subsampling on the RD curve. As the legend of Fig. 11(a) indicates, the considered videos were at full frame rate while the spatial resolutions were varied. The plot reveals the interesting tendency that lower spatial resolution videos are favored over the higher resolution videos as the target bit rate is reduced (increased compression). This is not unexpected, since retaining full video spatial dimension is not the best strategy at very low bit rates, as heavy compression artifacts are introduced. While spatial subsampling results in a loss of information and degradations of quality, much less compression is required to meet the bit budget, yielding less perceptually degraded videos. It is possible to construct a convex hull that can be used to help choose the best spatial subsampling strategy at each bit rate to maintain the best possible perceptual quality. We indicate such a spatial convex hull using a green dashed curve, in Fig. 11(b).

Fig. 11(b) also considers the effects of temporal subsampling. The videos were subsampled in both space and time. As shown in the figure, the perceptual quality at low bit rates can be further improved using simultaneous spatial and temporal subsampling, as compared to the convex hull constructed from just spatial subsampling. Fig. 11(c) depicts comprehensive RD curves obtained at original space-time resolutions, from the convex hull constructed using spatial subsampling only, and from the convex hull constructed using both space and time subsampling, along with 95% confidence intervals superimposed. It is easily observed that perceptual quality can be significantly improved at low bit rates by using space and/or time subsampling. Indeed, statistically superior quality improvements are obtained by considering both space and time subsampling, as opposed to considering only one kind of subsampling.

Of course, the aforementioned observations are comprehensive on the whole database, and the results may vary depending on the content characteristics. For example, Fig. 12 shows the RD curve of the 'American Football' sequence, which contains significant and rapid motions, including those arising from camera movements and from the action of football players. In this kind of video, the impact of temporal information loss can be much more significant than on more static contents. As shown in Fig. 12, the convex hull constructed from only spatial subsampling on this high-motion video produced better subjective quality results as compared to deploying any temporal subsampling. An optimal strategy for this content would likely employ only spatial subsampling, even at low bit rates. Understanding the effects of space-time subsampling as a function of content characteristics, and devising perceptually optimal strategies are among the topics that could be fruitfully investigated by analyzing the rich information available in the new database.

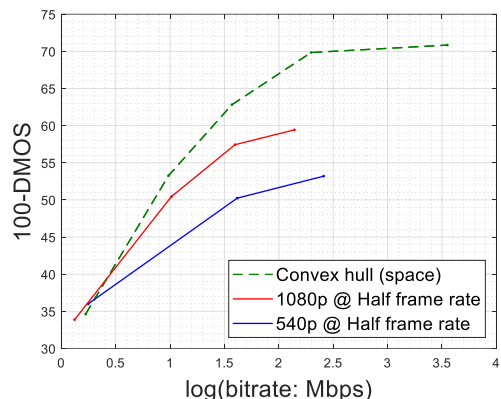

Fig. 12. Rate distortion curves of the video content 'American Football' at half frame rate, compared against the convex hull constructed using spatial subsampling.

## V. EVALUATION OF OBJECTIVE VIDEO QUALITY MODELS

As a way of both demonstrating and exploiting the usefulness of the new ETRI-LIVE STSVQ database we evaluated and compared the performances of a variety of relevant and widely-used image/video quality assessment models on it. We studied both Reference (including full and reduced reference) and No-reference models. The former assume that at least some information is derived from an undistorted reference image/video to compare against, while the latter predict image/video quality without using any reference information, which is often not available.

The reference models that we evaluated include the image quality assessment (IQA) models such as PSNR, SSIM [30], MSSSIM [31], and VIF [32], computed on each frame yielding predictions that were averaged (pooled) over all frames to obtain overall video quality scores. We also considered video

TABLE II

Performance comparison of VQA/IQA models on the ETRI-LIVE STSVQ database across different temporal subsampling levels. The two best models in each column are boldfaced.

| Model | | Full frame rate | | | | Half frame rate | | | | Overall (full + half frame rate) | | | |
|---|---|---|---|---|---|---|---|---|---|---|---|---|---|
| | | SRCC | KRCC | PLCC | RMSE | SRCC | KRCC | PLCC | RMSE | SRCC | KRCC | PLCC | RMSE |
| Reference | PSNR | 0.5736 | 0.4148 | 0.5652 | 13.09 | 0.3252 | 0.2298 | 0.2777 | 11.82 | 0.4179 | 0.2874 | 0.3955 | 12.49 |
| | SSIM [30] | 0.7582 | 0.5540 | 0.6790 | 14.16 | 0.3022 | 0.2044 | 0.1717 | 12.07 | 0.4953 | 0.3447 | 0.2435 | 13.19 |
| | MSSSIM [31] | 0.6472 | 0.4652 | 0.6116 | 13.65 | 0.2938 | 0.2081 | 0.2316 | 12.02 | 0.4431 | 0.3040 | 0.3187 | 12.89 |
| | VIF [32] | 0.6844 | 0.4902 | 0.6653 | 12.33 | **0.3932** | **0.2791** | 0.3281 | 11.81 | 0.5190 | 0.3577 | 0.4594 | 12.08 |
| | ST-RRED [33] | 0.7625 | 0.5544 | 0.6421 | 14.55 | 0.3212 | 0.2262 | 0.1781 | 11.88 | 0.4887 | 0.3395 | 0.1971 | 13.34 |
| | SpEED [34] | **0.8227** | **0.6181** | 0.6822 | 14.74 | 0.2480 | 0.1706 | 0.1541 | 11.86 | 0.4634 | 0.3272 | 0.1571 | 13.43 |
| | VMAF [35] | 0.7400 | 0.5648 | **0.7406** | **11.18** | **0.4995** | **0.3654** | **0.5002** | **10.90** | **0.5924** | **0.4311** | **0.5831** | **11.05** |
| | VSTR-ED [36] | **0.8840** | **0.7015** | **0.8770** | **11.92** | 0.3637 | 0.2500 | **0.3485** | 12.14 | **0.5358** | **0.3909** | 0.4679 | **12.03** |
| No Reference | BRISQUE [37] | 0.3868 | 0.2683 | 0.3713 | 14.49 | 0.3046 | 0.2175 | 0.2943 | **11.70** | 0.3462 | 0.2410 | 0.3422 | 13.23 |
| | NIQE [38] | 0.2869 | 0.2014 | 0.2612 | 15.09 | 0.2044 | 0.1380 | 0.1330 | 12.19 | 0.2476 | 0.1711 | 0.2067 | 13.77 |
| | TLVQM-HCF [39] | 0.2979 | 0.2707 | 0.2707 | 15.03 | 0.2470 | 0.1660 | 0.2265 | 11.92 | 0.2700 | 0.1821 | 0.2505 | 13.63 |

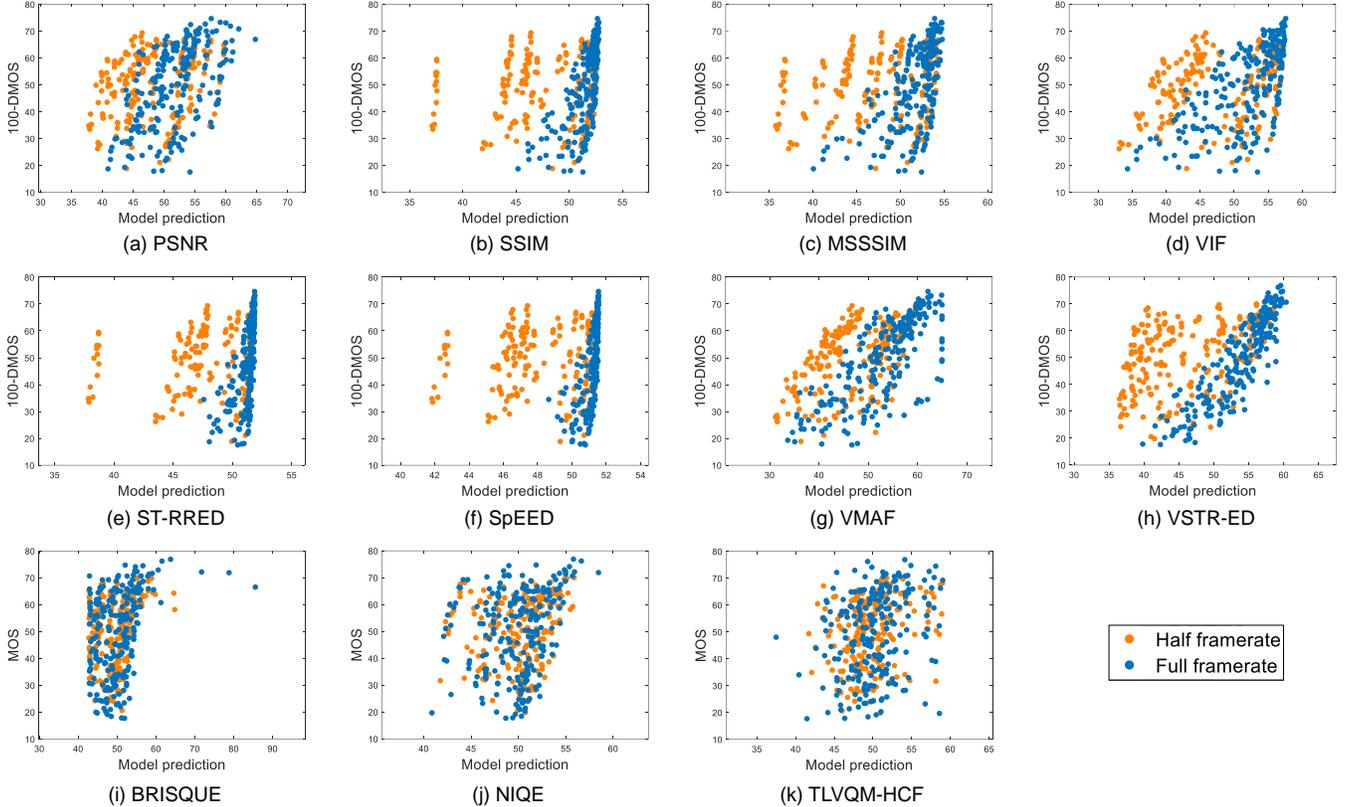

Fig. 13. Scatter plots of VQA/IQA model predictions plotted against subjective opinion scores of all of the distorted videos in the ETRI-LIVE STSVQ database. The orange data points refer to videos that were temporally subsampled to half frame rate, while the blue data points refer to full frame rate videos.

quality assessment (VQA) models that utilize both spatial and temporal video features. Among these, ST-RRED [33] and SpEED [34] are natural scene statistics (NSS) based models that measure the statistical space-time statistical deviations between a distorted videos and their references. VMAF [35] is a learning-based VQA model that fuses a set of quality-aware video features using a Support Vector Regressor (SVR). We also include a results of a recent prototype model we have developed [36], which is a space-time NSS based model that is based on statistical measurements of Video's Space-Time Regularity (VSTR). The prediction performance of the reference models is evaluated by comparing them to DMOS.

The no-reference models that we evaluated include BRISQUE [37] and NIQE [38], which are NSS-based, and TLVQM [39], which is a learning-based model that uses a large set of hand-designed features, including motion statistics and estimates of specific distortions. Since no-reference methods evaluate the intrinsic quality of videos, we evaluated no-reference model performance against MOS. MOS computation is similar to that for DMOS in Section IV-A, with (1) modified to read $d_{ijk} = s_{ijk}$.

The quality prediction performances of the compared models was evaluated using Spearman's rank order correlation coefficient (SRCC), the Kendall rank correlation coefficient (KRCC), the Pearson linear correlation coefficient (PLCC), and the root mean squared error (RMSE). SRCC and KRCC measure ordinal correlations, while PLCC measures linear correlation between variables. Higher values are favorable for SRCC, PLCC, and KRCC, and lower values are favorable for RMSE. Before computing PLCC and RMSE, we used logistic regression to linearize the model prediction following the procedure in [23].

| Model | | 540p | | 720p | | 1080p | | 2160p | | Overall | |
|---|---|---|---|---|---|---|---|---|---|---|---|
| | | SRCC | PLCC | SRCC | PLCC | SRCC | PLCC | SRCC | PLCC | SRCC | PLCC |
| Reference | PSNR | 0.4474 | 0.4270 | 0.3896 | 0.3667 | 0.3588 | 0.3215 | 0.4892 | 0.4588 | 0.4179 | 0.3955 |
| | SSIM [30] | 0.5116 | 0.2946 | 0.4601 | 0.2163 | 0.4411 | 0.1765 | 0.5831 | 0.2992 | 0.4953 | 0.2435 |
| | MSSSIM [31] | 0.4638 | 0.3849 | 0.4047 | 0.3088 | 0.3844 | 0.2512 | 0.5261 | 0.3574 | 0.4431 | 0.3187 |
| | VIF [32] | **0.5638** | **0.5248** | **0.5090** | **0.4427** | 0.4473 | 0.3787 | 0.5856 | 0.5066 | 0.5190 | 0.4594 |
| | ST-RRED [33] | 0.4698 | 0.2615 | 0.4254 | 0.1705 | 0.4385 | 0.1467 | 0.5876 | 0.2347 | 0.4887 | 0.1971 |
| | SpEED [34] | 0.4303 | 0.2083 | 0.4130 | 0.1273 | 0.4135 | 0.1075 | 0.5687 | 0.1990 | 0.4634 | 0.1571 |
| | VMAF [35] | **0.5223** | **0.5206** | **0.5607** | **0.5541** | **0.5400** | **0.5204** | **0.6916** | **0.6713** | **0.5924** | **0.5831** |
| | VSTR-ED [36] | 0.4835 | 0.4260 | 0.4850 | 0.3985 | **0.4875** | **0.3951** | **0.6461** | **0.5821** | **0.5358** | **0.4679** |
| No Reference | BRISQUE [37] | 0.2350 | 0.1964 | 0.2740 | 0.2446 | 0.3146 | 0.3046 | 0.4895 | 0.4575 | 0.3462 | 0.3422 |
| | NIQE [38] | 0.2094 | 0.0844 | 0.1343 | 0.0357 | 0.2060 | 0.1377 | 0.3768 | 0.3704 | 0.2476 | 0.2067 |
| | TLVQM-HCF [39] | 0.1179 | 0.1310 | 0.2599 | 0.2585 | 0.3447 | 0.3002 | 0.3156 | 0.3070 | 0.2700 | 0.2505 |

| | PSNR | SSIM | MSSSIM | VIF | ST-RRED | SpEED | VMAF | VSTR-ED | BRISQUE | NIQE | TLVQM-HCF |
|---|---|---|---|---|---|---|---|---|---|---|---|
| PSNR | ------- | ------- | ------- | ------- | ------- | ------- | -0---0 | -0----- | ------- | -1----1 | -1----- |
| SSIM | ------- | ------- | ------- | -0----- | ------- | ------- | 00---00 | -0----- | ------- | ------- | ------- |
| MSSSIM | ------- | ------- | ------- | ------- | ------- | ------- | -0---00 | -0----- | ------- | ------- | ------- |
| VIF | ------- | -1----- | ------- | ------- | -1---1 | -1---1 | ------- | ------- | -1----- | -1---1 | -1---1 |
| ST-RRED | ------- | ------- | ------- | -0---0 | ------- | ------- | -0---00 | -0---0 | ------- | ------- | ------- |
| SpEED | ------- | ------- | ------- | -0---0 | ------- | ------- | -0---00 | -0---0 | ------- | ------- | ------- |
| VMAF | -1---1 | 11---11 | -1---11 | ------- | -1---11 | -1---11 | ------- | ------- | -1---11 | 11-1-11 | 11---11 |
| VSTR-ED | -1----- | -1----- | -1----- | ------- | -1---1 | -1---1 | ------- | ------- | -1---1 | -1---1 | -1---11 |
| BRISQUE | ------- | ------- | ------- | -0----- | ------- | ------- | -0---0 | -0----- | ------- | ------- | ------- |
| NIQE | -0---0 | ------- | ------- | -0---0 | ------- | ------- | 00-0-00 | -0---00 | ------- | ------- | ------- |
| TLVQM-HCF | -0----- | ------- | ------- | -0---0 | ------- | ------- | -0---00 | ------- | ------- | ------- | ------- |

## A. Overall Performance Comparison

We evaluated the prediction performance of the compared models on the new ETRI-LIVE STSVQ database. For models that involve machine learning, we first tested the model as pre-trained on another database, or used one of their representative features. in this way, we measured generalized model performance over all 437 videos of our database without applying train-test split procedures. Of course, we also obtained cross-validation results of all learned models trained on our database, in the following section. For VMAF, we used the vmaf_4k_v.0.6.1 model available at [40]. For VSTR, we used one of its representative features, an entropic difference (ED) feature computed on space-time displaced frame differences. For BRISQUE, we used the model trained on the LIVE image database provided by the authors. For TLVQM, we used the average of thirty high complexity features (HCF) to capture a wide variety of distortions.

Table II reports the performances of the compared models over different temporal subsampling levels and overall. Fig. 13 shows scatter plots of the model predictions against the subjective opinion scores. As shown in the Table, the no-reference models performed much worse than the reference models. It is interesting that the references models attained very high performance on full frame rate videos, which they have been amply validated on in the past, but performed much worse on temporally subsampled videos. This suggests that there is ample room for improvement of solutions to the underdeveloped topic of assessing the quality of temporally subsampled and compressed videos. Overall, VIF, VMAF and

VSTR-ED yielded the highest prediction performances.

Table III tabulates the prediction performance against amount of spatial subsampling. The no-reference models again underperformed against the reference models. However, unlike the results in Table II, similar prediction performance was obtained across spatial resolutions. VMAF yielded good performance across all resolutions, VIF delivered good performance at lower spatial resolutions (540p and 720p), and VSTR-ED delivered good performance at higher spatial resolutions (1080p and 2160p).

## B. Statistical Significance

We verified the statistical significance of the performance differences among the compared models in Tables II and III via an F-test. Table IV shows the F-test results performed on the residuals between the model predictions and the subjective opinion scores. The underlying assumption is that the distribution of residuals follows a zero mean Gaussian distribution. The F-test evaluates the ratio of variances of the residuals, and determines whether the variances are equal at the 95% confidence level. Table IV contains 7 entries, corresponding to half frame rate, full frame rate, 540p, 720p, 1080p, 2160p, and all of the videos, in that order. As shown in the Table, VIF, VMAF and VSTR-ED attained statistically superior prediction performances as compared to the other methods.

## C. Cross-validation Results on Learning-based Models

We also evaluated the cross-validation performances of the learning-based models trained specifically on our database. The trained models include VMAF, VSTR, BRISQUE, TLVQM, and VIDEVAL [42] which each consists of 6, 16, 36, 75, and

TABLE V
CROSS-VALIDATION PERFORMANCE COMPARISON OF VQA/IQA MODELS ON THE ETRI-LIVE STSVQ DATABASE ACROSS DIFFERENT TEMPORAL SUBSAMPLING LEVELS. THE NUMBERS DENOTE MEDIAN VALUES FOR 1000 ITERATION OF RANDOMLY SPLIT TRAIN AND TEST SETS. THE VALUES INSIDE THE BRACKETS DENOTE STANDARD DEVIATION. THE TWO BEST MODELS IN EACH COLUMN ARE BOLDFACED.

| Model | | Full frame rate | | Half frame rate | | Overall (full + half frame rate) | |
|---|---|---|---|---|---|---|---|
| | | SRCC | PLCC | SRCC | PLCC | SRCC | PLCC |
| Reference | PSNR | 0.6791 (0.1542) | 0.6535 (0.1471) | 0.5079 (0.1803) | 0.4021 (0.2186) | 0.5170 (0.1333) | 0.4757 (0.1501) |
| | SSIM [30] | 0.8143 (0.0887) | 0.7596 (0.1380) | 0.4708 (0.1919) | 0.2673 (0.2255) | 0.5348 (0.1407) | 0.3346 (0.2210) |
| | MSSSIM [31] | 0.7539 (0.1365) | 0.7061 (0.1416) | 0.4726 (0.1809) | 0.3159 (0.2132) | 0.5295 (0.1318) | 0.3812 (0.1927) |
| | VIF [32] | 0.7588 (0.1183) | 0.7338 (0.1378) | 0.5287 (0.1781) | 0.4617 (0.2444) | 0.6022 (0.1316) | 0.5412 (0.1730) |
| | ST-RRED [33] | 0.8346 (0.1162) | 0.7350 (0.1650) | 0.4742 (0.1931) | 0.2738 (0.2251) | 0.5220 (0.1482) | 0.2705 (0.1996) |
| | SpEED [34] | **0.8660 (0.0799)** | **0.7656 (0.1412)** | 0.4424 (0.1883) | 0.2549 (0.1974) | 0.4898 (0.1238) | 0.2205 (0.1685) |
| | VMAF [35] | 0.7812 (0.1553) | 0.7644 (0.1513) | **0.6213 (0.2022)** | **0.5973 (0.2261)** | **0.6717 (0.1586)** | **0.6580 (0.1672)** |
| | VSTR [36] | **0.8921 (0.0716)** | **0.8915 (0.0771)** | **0.6359 (0.1718)** | **0.6227 (0.1784)** | **0.7625 (0.0946)** | **0.7563 (0.0962)** |
| No Reference | BRISQUE [37] | 0.4039 (0.2243) | 0.3747 (0.2133) | 0.3182 (0.2260) | 0.2999 (0.2157) | 0.3517 (0.2245) | 0.3414 (0.2122) |
| | NIQE [38] | 0.4025 (0.1957) | 0.4175 (0.2172) | 0.3545 (0.2044) | 0.3242 (0.2170) | 0.3728 (0.1977) | 0.3751 (0.2199) |
| | TLVQM [39] | 0.4928 (0.1774) | 0.4813 (0.1840) | 0.3623 (0.1870) | 0.3472 (0.1936) | 0.4261 (0.1726) | 0.4234 (0.1754) |
| | VIDEVAL [42] | 0.4188 (0.1927) | 0.3869 (0.1960) | 0.3306 (0.1986) | 0.3037 (0.2047) | 0.3601 (0.1922) | 0.3323 (0.1976) |

TABLE VI
CROSS-VALIDATION PERFORMANCE COMPARISON OF VQA/IQA MODELS ON THE ETRI-LIVE STSVQ DATABASE ACROSS DIFFERENT SPATIAL SUBSAMPLING LEVELS. THE NUMBERS DENOTE MEDIAN VALUES FOR 1000 ITERATION OF RANDOMLY SPLIT TRAIN AND TEST SETS. THE VALUES INSIDE THE BRACKETS DENOTE STANDARD DEVIATION. THE TWO BEST MODELS IN EACH COLUMN ARE BOLDFACED.

| Model | | 540p | | 720p | | 1080p | | 2160p | | Overall | |
|---|---|---|---|---|---|---|---|---|---|---|---|
| | | SRCC | PLCC | SRCC | PLCC | SRCC | PLCC | SRCC | PLCC | SRCC | PLCC |
| Reference | PSNR | 0.51 (0.17) | 0.47 (0.19) | 0.48 (0.18) | 0.44 (0.20) | 0.48 (0.17) | 0.41 (0.18) | 0.59 (0.13) | 0.53 (0.14) | 0.52 (0.13) | 0.48 (0.15) |
| | SSIM [30] | 0.54 (0.18) | 0.37 (0.23) | 0.50 (0.16) | 0.29 (0.23) | 0.50 (0.17) | 0.25 (0.23) | 0.64 (0.13) | 0.40 (0.21) | 0.53 (0.14) | 0.33 (0.22) |
| | MSSSIM [31] | 0.53 (0.17) | 0.42 (0.21) | 0.49 (0.17) | 0.36 (0.22) | 0.47 (0.16) | 0.31 (0.20) | 0.61 (0.12) | 0.43 (0.18) | 0.53 (0.13) | 0.38 (0.19) |
| | VIF [32] | **0.63 (0.17)** | **0.59 (0.20)** | 0.60 (0.17) | 0.53 (0.22) | 0.56 (0.16) | 0.47 (0.21) | 0.68 (0.12) | 0.58 (0.16) | 0.60 (0.13) | 0.54 (0.17) |
| | ST-RRED [33] | 0.52 (0.18) | 0.31 (0.19) | 0.47 (0.16) | 0.25 (0.20) | 0.49 (0.17) | 0.23 (0.21) | 0.63 (0.14) | 0.36 (0.21) | 0.52 (0.15) | 0.27 (0.20) |
| | SpEED [34] | 0.46 (0.17) | 0.26 (0.17) | 0.43 (0.15) | 0.17 (0.17) | 0.45 (0.16) | 0.17 (0.17) | 0.60 (0.10) | 0.27 (0.17) | 0.49 (0.12) | 0.22 (0.17) |
| | VMAF [35] | 0.59 (0.21) | 0.59 (0.23) | **0.64 (0.19)** | **0.62 (0.21)** | **0.64 (0.18)** | **0.61 (0.19)** | **0.75 (0.14)** | **0.73 (0.14)** | **0.67 (0.16)** | **0.66 (0.17)** |
| | VSTR [36] | **0.72 (0.13)** | **0.74 (0.13)** | **0.74 (0.12)** | **0.73 (0.12)** | **0.74 (0.11)** | **0.72 (0.11)** | **0.83 (0.08)** | **0.81 (0.09)** | **0.76 (0.09)** | **0.76 (0.10)** |
| No Reference | BRISQUE [37] | 0.33 (0.22) | 0.30 (0.22) | 0.30 (0.24) | 0.27 (0.23) | 0.39 (0.25) | 0.35 (0.25) | 0.50 (0.24) | 0.49 (0.21) | 0.35 (0.22) | 0.34 (0.21) |
| | NIQE [38] | 0.35 (0.25) | 0.27 (0.25) | 0.34 (0.22) | 0.28 (0.24) | 0.39 (0.24) | 0.34 (0.25) | 0.44 (0.21) | 0.45 (0.22) | 0.37 (0.20) | 0.38 (0.22) |
| | TLVQM [39] | 0.26 (0.16) | 0.26 (0.17) | 0.36 (0.20) | 0.35 (0.21) | 0.45 (0.20) | 0.43 (0.20) | 0.60 (0.15) | 0.58 (0.15) | 0.43 (0.17) | 0.42 (0.18) |
| | VIDEVAL [42] | 0.24 (0.18) | 0.19 (0.18) | 0.23 (0.22) | 0.21 (0.21) | 0.37 (0.23) | 0.31 (0.24) | 0.59 (0.19) | 0.56 (0.18) | 0.36 (0.19) | 0.33 (0.20) |

60 features, respectively. The features from the models were used to train an SVR with a radial basis function (RBF) kernel. The SVR-RBF parameters were determined using cross-validation within the training set, as in [41].

Since the ETRI-LIVE database consist of videos afflicted by various distortions applied on the same source contents, we took particular care to separate the train and test sets 'content-wise.' Hence, the train and test sets did not share any videos derived from the same source contents. For the performance evaluation, we used 5-fold cross validation. Since the database contains 15 unique source contents, the model was trained on all the distorted versions of 12 source contents (and their DMOS or MOS, as appropriate), and tested on the distorted videos derived from the other three source contents. We ran 1000 train/test iterations, in this manner, where the train/test sets were randomly divided at each iteration while following the content-wise separation. The results in Table V and VI show the medians and standard deviations of PLCC and SRCC of the compared learning-based models across the 1000 iterations. We also list the median performances of the other non-learning-based models, but on the *same* randomized splits for comparison.

Table V shows the cross-validation performance over different temporal subsampling levels and overall. No-reference models underperformed against the reference models. Reference models again attained high performances on full frame rate videos, but performed comparatively worse on temporally subsampled videos. Overall, the learning-based reference models, VMAF and VSTR, outperformed other models.

Table VI presents the cross-validation performance over different spatial subsampling level and overall. As seen in Table III, similar prediction performance was obtained across spatial resolutions. Except at the lowest resolution (540p), the learning-based reference models again outperformed other models.

## VI. CONCLUSION AND FUTURE WORK

We conducted a large-scale human study to more generally understand combined space-time subsampling and compression affect the perceptual quality of videos. The new ETRI-LIVE STSVQ database contains 15 unique source contents and 437 distorted versions of them, on which almost 15,000 subjective opinion scores were collected. This study is the first to include the subjective scores on 4K 10bit videos with frame rates up to 120Hz, subjected to simultaneous space-time subsampling and compression.

Analysis of the subjective scores reveals that, while space-time subsampling inevitably results in a loss of information and subsequent degradations on quality, it may be a good tradeoff against increased compression, given a fixed bit rate budget. A rate distortion analysis of the subjective scores showed that space-time subsampling prior to video compression can significantly improve video perceptual quality at low bit rates. An interesting topic for further study, would be to understand content-wise trade-offs between space-time subsampling and compression, perhaps leading to more optimal space-time semantic resolution adaptation strategies for perceptual video coding.

We evaluated several high-performance image/video quality models on our database. The results from this benchmark study indicate that, while the compared models can effectively predict the quality of videos subjected to spatial subsampling and compression, they are much less effective if temporal subsampling is included in the mix of distortions. This suggests that further study could lead to significant improvements of existing models, or new models altogether, more capable of capturing the deleterious perceptual effects of temporal subsampling.

ACKNOWLEDGMENT

This work was supported by an Institute for Information & Communications Technology Promotion (IITP) grant funded by the Korean government (MSIT) (No. 2017-0-00072, Development of Audio/Video Coding and Light Field Media Fundamental Technologies for Ultra Realistic Teramedia).